\documentclass[twocolumn,a4]{jpsj3}
\def\runauthor{D. Aoki \textit{et al.}}

\title{Fermi Surface Instabilities in Ferromagnetic Superconductor URhGe}

\author{
Dai~{\sc Aoki}$^{1,2}$\thanks{E-mail: aoki@imr.tohoku.ac.jp}, 
Georg~{\sc Knebel}$^2$,
and
Jacques~{\sc Flouquet}$^2$
}

\inst{%
$^1$IMR, Tohoku University, Oarai, Ibaraki, 311-1313, Japan\\
$^2$INAC/SPSMS, CEA-Grenoble, 38054 Grenoble, France
}

\abst{%
The field-reentrant (field-reinforced) superconductivity on ferromagnetic superconductors is one of the most interesting topics in unconventional superconductivity.
The enhancement of effective mass and the induced ferromagnetic fluctuations play key roles for reentrant superconductivity.
However, the associated change of the Fermi surface, which is often observed at (pseudo-) metamagnetic transition,
can also be a key ingredient.
In order to study the Fermi surface instability, we performed Hall effect measurements in the ferromagnetic superconductor URhGe.
The Hall effect of URhGe is well explained by two contributions, namely by the normal Hall effect and by the large anomalous Hall effect due to skew scattering.
The large change in the Hall coefficient is observed at low fields between the paramagnetic and ferromagnetic states for $H\parallel c$-axis (easy-magnetization axis) in the orthorhombic structure,
indicating that the Fermi surface is reconstructed in the ferromagnetic state below the Curie temperature ($T_{\rm Curie}=9.5\,{\rm K}$).
At low temperatures ($T \ll T_{\rm Curie}$),
when the field is applied along the $b$-axis, the reentrant superconductivity was observed in both the Hall resistivity and the magnetoresistance below $0.4\,{\rm K}$.
Above $0.4\,{\rm K}$, a large jump with the first-order nature was detected in the Hall resistivity at a spin-reorientation field $H_{\rm R}\sim 12.5\,{\rm T}$,
demonstrating that the marked change of the Fermi surface occurs between the ferromagnetic state and the polarized state above $H_{\rm R}$.
The results can be understood by the Lifshitz-type transition, induced by the magnetic field or by the change of the effective magnetic field.
}

\begin{document}
\maketitle
\section{Introduction}
The coexistence of ferromagnetism (FM) and superconductivity (SC) attracts much interest
because unconventional superconductivity is expected.~\cite{Aok12_JPSJ_review}
In the conventional view, SC competes against FM,
since the strong internal field due to the FM order
easily destroys Cooper pairs.

The first case of the microscopic coexistence of FM and SC was found in UGe$_2$,~\cite{Sax00}
where SC appears in the FM phase just below the FM critical pressure $P_{\rm c}\sim 1.5\,{\rm GPa}$,
where the FM state changes into the paramagnetic (PM) state.
After the discovery of UGe$_2$, two other uranium ferromagnets, URhGe and UCoGe, were found to 
be superconductors even at ambient pressure.~\cite{Aok01,Huy07}
It is considered that the triplet state of Cooper pairs is responsible for SC,
because it can survive even in the strong internal field due to FM.
One of the most spectacular characteristics is
the field-reinforced (field-reentrant) superconductivity (RSC).~\cite{Lev05,Aok09_UCoGe}
In URhGe and UCoGe, when the field is applied along the hard-magnetization axis ($b$-axis),
the FM Curie temperature $T_{\rm Curie}$ is suppressed to $0\,{\rm K}$.
The simple image is that
the effective mass of conduction electrons increases in the region of $T_{\rm Curie}\to 0$ and then
SC is reinforced under a magnetic field.
Our previous results of resistivity, magnetization, and specific heat measurements clearly indicate that
the resistivity $A$ coefficient and the specific heat $\gamma$-value increase at high fields for $H\parallel b$-axis,
whereas they decrease for $H\parallel c$-axis (easy-magnetization axis).~\cite{Miy08,Har11,Aok11_ICHE,Aok09_UCoGe}
Correspondingly, the suppression of the FM longitudinal fluctuation is observed in UCoGe by NMR experiments
when the field is applied along the easy-magnetization axis ($c$-axis).~\cite{Hat12}

Up to now, the Fermi surface has been assumed to be unchanged under a magnetic field.
In reality, the Fermi surface can be affected by the magnetic field owing to 
the polarization of the moment between the majority- and minority-spin bands,
or by the change of the ground state itself.
In the triplet equal-spin pairing,
$H_{\rm c2}$, which is governed by the orbital limit, is linked to the Fermi velocity $v_{\rm F}$ by the relation $H_{\rm c2}\propto (T_{\rm sc}/v_{\rm F})^2$.
Thus, the enhancement of $H_{\rm c2}$ can be induced by ether the collapse of the Fermi wave vector $k_{\rm F}$
or the enhancement of the effective mass $m^\ast$.
Therefore, it is important to clarify the interplay between the Fermi surface instability and the superconductivity.

A clear example is UGe$_2$, in which the Fermi surfaces markedly change among FM1 (weakly polarized phase), FM2 (strongly polarized phase), and PM, as detected by de Haas-van Alphen (dHvA) experiments.~\cite{Ter01,Set02} 
Corresponding to the change of Fermi surfaces, the $H_{\rm c2}$ curve shows an S-shape at a pressure of $P_{\rm x} < P <P_{\rm c}$,
where $P_{\rm x}$ is the critical pressure between FM2 and FM1, and $P_{\rm c}$ is that between FM1 and PM.~\cite{She01}
In UCoGe, the S-shaped $H_{\rm c2}$ curve for $H\parallel b$-axis was qualitatively explained by the
results of thermopower measurement, which is a sensitive probe for the Fermi surface change~\cite{Mal12}.
Furthermore, the Shubnikov-de Haas (SdH) experiments show the modification of the Fermi surface at high fields above $20\,{\rm T}$,
indicating that UCoGe is a low carrier system associated with a large effective mass,~\cite{Aok11_UCoGe}
which is favorable for the field-induced Fermi surface change.
In URhGe, the SdH experiments reveal the collapse of a small pocket Fermi surface around the spin reorientation field $H_{\rm R}$.~\cite{Yel11}
However, in both UCoGe and URhGe, the Fermi surface is not fully determined experimentally because of the insufficient sample quality and heavy effective mass.

Thus, in order to study the RSC and Fermi surface instabilities,
we have chosen URhGe and measured the Hall effect at low temperatures at high fields with the precise tuning of field directions,
using high-quality single crystals.

\begin{figure}[tbh]
\begin{center}
\includegraphics[width=1 \hsize,clip]{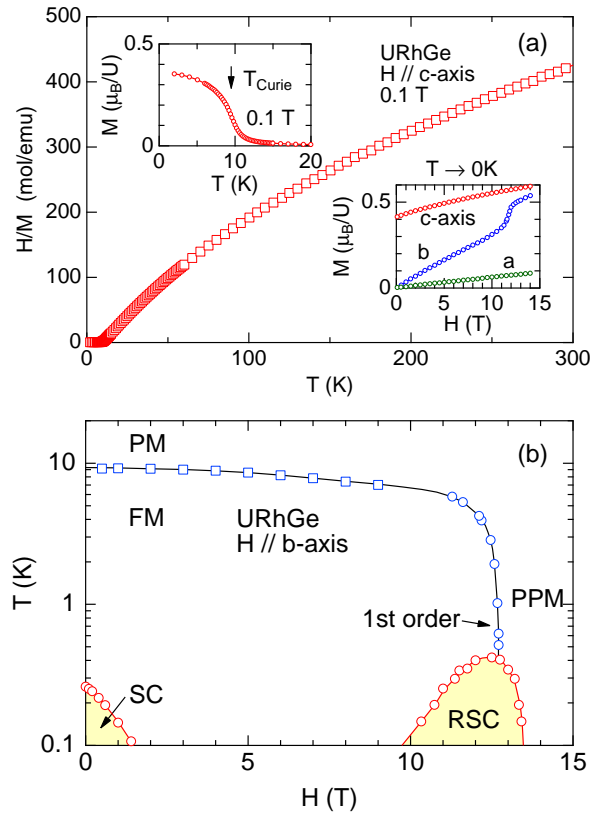}
\end{center}
\caption{(Color online) (a) Temperature dependence of the inverse susceptibility at $0.1\,{\rm T}$ for $H\parallel c$-axis in URhGe. The top-left inset shows the temperature dependence of the magnetization at low temperatures after field-cooling. The bottom-right inset shows the magnetization curves for $H\parallel a$-, $b$- and $c$-axes extrapolated to $0\,{\rm K}$ cited from Ref.~\citen{Har11}. (b) Temperature-field phase diagram for $H\parallel b$-axis.~\cite{Aok12_JPSJ_review}
PM, FM and PPM denote paramagnetism, ferromagnetism and polarized paramagnetism, respectively.
SC and RSC denote the superconductivity and reentrant superconductivity, respectively.}
\label{fig:Mag_phase}
\end{figure}

URhGe crystallizes in the TiNiSi-type orthorhombic crystal structure.
The FM order occurs at $T_{\rm Curie}=9.5\,{\rm K}$ with the ordered moment of $0.42\,\mu_{\rm B}/{\rm U}$,
as shown in Fig.~\ref{fig:Mag_phase}(a).
The moment is directed along the $c$-axis with a collinear structure.
The SC appears below $T_{\rm sc}=0.25\,{\rm K}$ at zero field.
The electronic specific heat coefficient is $160\,{\rm mJ\, K^{-2} mol^{-1}}$, 
indicating that URhGe is a moderately enhanced heavy fermion system.
When the field is applied along the $b$-axis (hard-magnetization axis),
the moment starts to tilt from the $c$-axis to the $b$-axis with the field, and 
finally the moment is completely directed along the $b$-axis above the spin reorientation field $H_{\rm R}\sim 12\,{\rm T}$,
which is connected to the decrease in $T_{\rm Curie}$ under a magnetic field.
Interestingly, the RSC appears around the field window approximately from $9$ to $13\,{\rm T}$.
The temperature-field phase diagram~\cite{Aok12_JPSJ_review} and magnetization curves~\cite{Har11} are shown in Fig.~\ref{fig:Mag_phase}(b) and the bottom-right inset of Fig.~\ref{fig:Mag_phase}(a), respectively.

In this paper, we present the results of Hall effect measurements in URhGe for $H \parallel a$-, $b$-, and $c$-axes.
From the low-field measurements for $H\parallel c$-axis, it is found that the Hall coefficient in the FM state changes from that in the PM state, indicating the reconstruction of the Fermi surface below $T_{\rm Curie}$.
A large jump of Hall resistivity at $H_{\rm R}$ for $H\parallel b$-axis implies that the Fermi surfaces markedly change also through $H_{\rm R}$.
The first-order transition at $H_{\rm R}$ for $H\parallel b$-axis was clearly detected in Hall resistivity with a hysteresis,
which is immediately suppressed by tilting the field direction slightly to the $c$-axis.
In addition, an anisotropic response of magnetoresistance for $H\parallel b$-axis between $J \parallel a$-, $b$- $c$-axes is found.
These results suggest that two effects are favorable for RSC in URhGe.
One is the ferromagnetic fluctuation, which was already observed as the enhancement of the effective mass $m^{\ast}$.
The other is the Fermi surface instability, which is detected by the present Hall effect measurements and previous SdH experiments~\cite{Yel11}.

\section{Experimental}
High-quality single crystals of URhGe were grown by the Czochralski method in a tetra-arc furnace. 
The grown single crystals were annealed under ultra high vacuum at high temperatures.
The single-crystal ingot was then oriented by taking X-ray Laue photographs and cut using a spark cutter.
The quality of the single crystals was checked by resistivity measurements at low temperatures down to $0.1\,{\rm K}$ using a homemade adiabatic demagnetization refrigerator (ADR) combined with a commercial PPMS.
All samples in the present studies show superconductivity at $\sim 0.25\,{\rm K}$,
and RSC was confirmed for $H\parallel b$-axis.
The residual resistivity ratio (RRR) is 20--40.
The thin samples for Hall effect measurements with a thickness of $0.15$--$0.05\,{\rm mm}$ were 
prepared for $H\parallel a$-, $b$- and, $c$-axes. 
The Hall effect was measured by the four-probe AC method ($f\sim 17\,{\rm Hz}$) at high fields up to $16\,{\rm T}$ and at low temperatures down to $0.1\,{\rm K}$.
The field was applied for both positive and negative directions to eliminate the effect of magnetoresistance.
In addition, the magetoresistance was measured by the four-probe AC method under the same experimental conditions using the same samples.
The magnetization and susceptibility, which were used for the analysis of Hall effect measurements, were measured by a commercial SQUID magnetometer at temperature down to $2\,{\rm K}$ and at high fields up to $5.5\,{\rm T}$.
For the analysis of high-field Hall effect data for $H\parallel b$-axis at low temperatures,
the magnetization data in Ref.~\citen{Har11} were used.

\section{Results and Discussion}
\begin{figure}[tbh]
\begin{center}
\includegraphics[width=1 \hsize,clip]{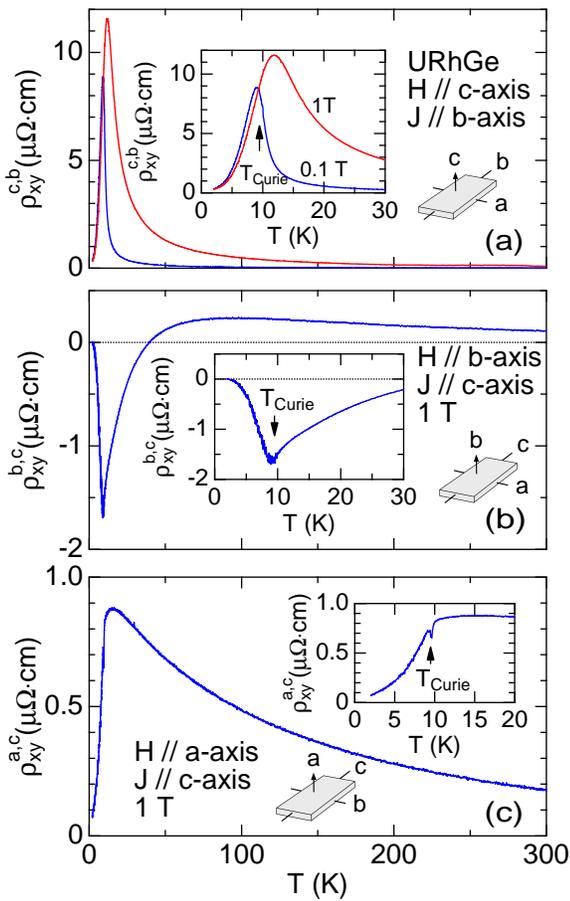}
\end{center}
\caption{(Color online) Temperature dependence of Hall resistivity in URhGe. 
(a) Hall resistivity for $H\parallel c$-axis at $0.1$ and $1\,{\rm T}$ for the current along $b$-axis ($J\parallel b$-axis). 
(b) Hall resistivity for $H\parallel b$-axis and $J\parallel c$-axes at $1\,{\rm T}$.
(c) Hall resistivity for $H\parallel a$-axis and $J\parallel c$-axis at $1\,{\rm T}$.
The insets show the Hall resistivity at low temperatures. 
The measurement configurations are illustrated in each panel.
}
\label{fig:Hall_Tdep}
\end{figure}

Figure~\ref{fig:Hall_Tdep} shows the temperature dependences of the Hall resistivity for $H\parallel a$- and $b$-, and $c$-axes in URhGe.
As shown in Fig.~\ref{fig:Hall_Tdep}(a),
when a small field ($0.1\,{\rm T}$) is applied along the easy-magnetization axis ($H\parallel c$-axis),
the Hall resistivity increases on cooling from room temperature and shows a peak just below $T_{\rm Curie}$ ($=9.5\,{\rm K}$).
By applying a higher field ($1\,{\rm T}$), a broad and larger maximum is observed at approximately $12\,{\rm K}$.
These results display typical behaviors of the Hall effect in ferromagnets.
The Hall resistivity $\rho_{xy}$ can be described by
\begin{equation}
\rho_{xy} = R_{\rm 0} H + R_{\rm s} M,
\label{eq1}
\end{equation} 
where $R_0$ is the normal Hall coefficient and the second term is attributed to the anomalous Hall effect with the magnetization $M$.
The anomalous Hall effect originates from skew scattering, side jump scattering, and the Berry phase. 
In general, the anomalous Hall effect is very large in ferromagnets.
The decrease in $\rho_{xy}$ with decreasing temperature below $9\,{\rm K}$ is mainly due to the 
strong decrease in resistivity in the FM state,
which plays an important role in the anomalous Hall effect.

\begin{figure}[tbh]
\begin{center}
\includegraphics[width=1 \hsize,clip]{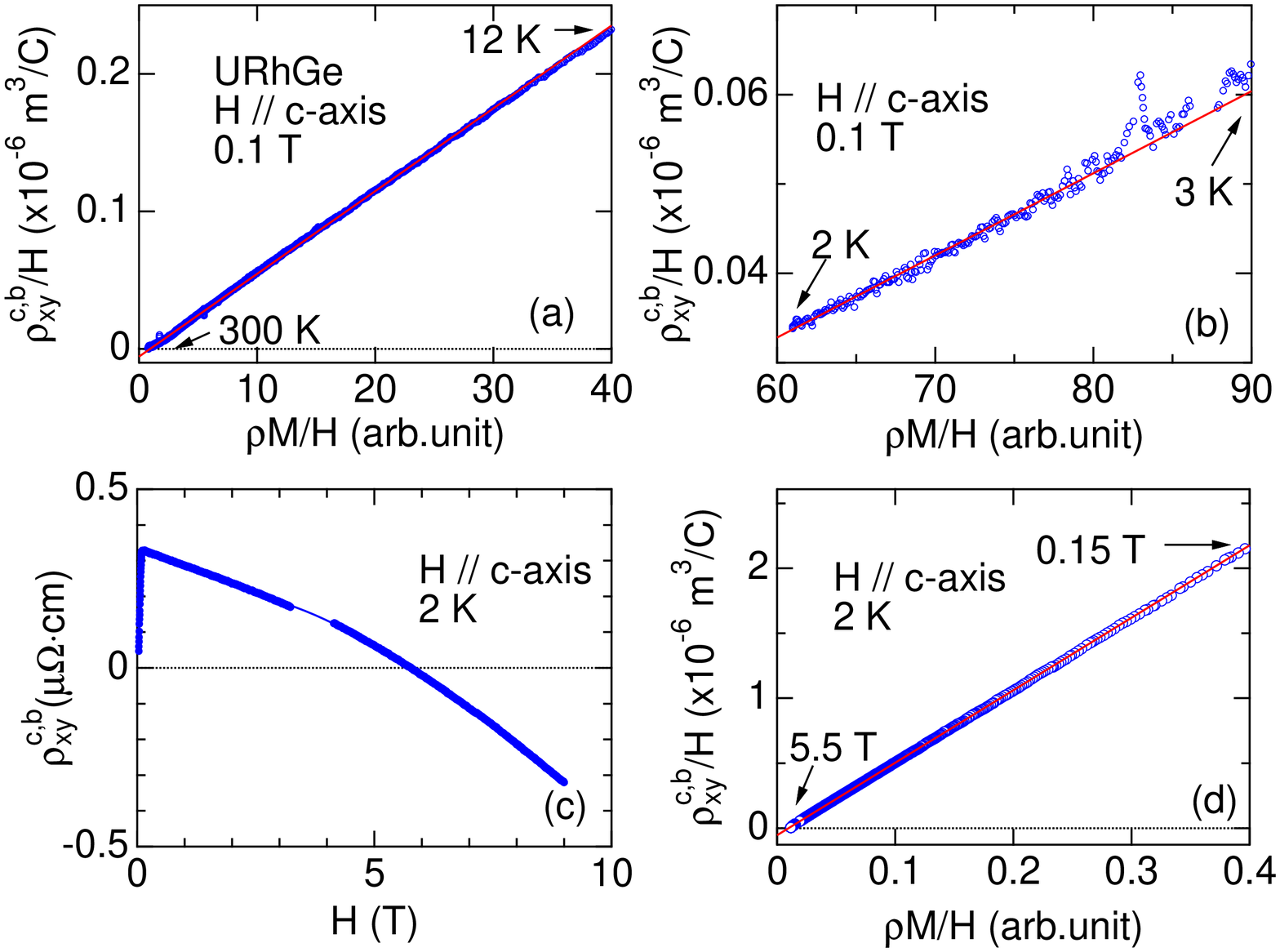}
\end{center}
\caption{(Color online) (a) Hall resistivity for $H\parallel c$-axis of URhGe at $0.1\,{\rm T}$ in the PM state obtained by temperature sweep in the form of $\rho_{xy}/H$ vs $\rho M/H$. (b) Hall resistivity at $0.1\,{\rm T}$ in the FM state obtained by temperature sweep in the form of $\rho_{xy}/H$ vs $\rho M/H$.  (c) Field dependence of the Hall resistivity at $2\,{\rm K}$. (d) Hall resistivity at $2\,{\rm K}$ in the form of $\rho_{xy}/H$ vs $\rho M/H$ obtained by field sweep.}
\label{fig:Hall_analysis}
\end{figure}

To extract the Hall coefficient $R_0$, 
the Hall resistivity data are plotted in the form of $\rho_{xy}/H$ vs $\rho M/H$,
assuming that the anomalous Hall effect mainly originates from skew scattering, 
namely, $R_{\rm s}\propto \rho$, as shown in Fig.~\ref{fig:Hall_analysis}(a).
A good linear relation is found in the wide temperature range from $300$ to $12\,{\rm K}$.
The intercept for $\rho M/H \to 0$ gives $R_0=-5.6\times 10^{-9}\,{\rm m^3/C}$ in the 
paramagnetic state.
Here, we assume that the normal Hall coefficient is constant above $T_{\rm Curie}$.
Assuming the single-band model with $|R_0|=1/(ne)$, we obtain the carrier number $n=1.1\times 10^{27}\,{\rm /m^3}$, which is equal to $0.25$ electrons/unit cell.
The negative sign of $R_0$ indicates that the carrier is dominated by electrons.

It should be noted that $\rho_{xy}$ in Fig.~\ref{fig:Hall_Tdep}(a) is always positive
because of the large positive contribution of the anomalous Hall effect, plus the small negative contribution of the normal Hall effect.
For example, at $100\,{\rm K}$ at $0.1\,{\rm T}$,
the contribution of the normal Hall effect is $R_0 H = -0.056 \,\mu\Omega\!\cdot\!{\rm cm}$,
while the anomalous Hall effect will give $R_{\rm s}M=0.104\,\mu\Omega\!\cdot \!{\rm cm}$.
Thus, the Hall resistivity in total has a positive sign with $\rho_{xy}=0.10\,\mu\Omega\!\cdot\!{\rm cm}$.

In the FM state well below $T_{\rm Curie}$,
a good linear relation was also found between $2$ and $3\,{\rm K}$, as shown in Fig.~\ref{fig:Hall_analysis}(b),
which gives $R_0=-2.2\times 10^{-8}\,{\rm m^3/C}$.
Since the linear fit as a function of $\rho M/H$ is only limited, 
the field dependence of $\rho_{xy}$ was also measured at $2\,{\rm K}$, as shown in Fig.~\ref{fig:Hall_analysis}(c).
Following the same method mentioned above, a good linear relation is again obtained,
as shown in Fig.~\ref{fig:Hall_analysis}(d).
The obtained $R_0$ is $-5.4\times 10^{-8}\,{\rm m^3/C}$ at $2\,{\rm K}$ 
in the ferromagnetic state.
This value is not very far from that obtained from the temperature scan in Fig.~\ref{fig:Hall_analysis}(b),
supporting the validity of the fitting.
The large change in $R_0$ between the PM and FM states with a one order magnitude difference 
implies that the Fermi surface is reconstructed at the FM transition.

For $H\parallel b$-axis, which corresponds to the hard-magnetization axis,
a sharp kink is observed at $T_{\rm Curie}$,
as shown in Fig.~\ref{fig:Hall_Tdep}(b).
At room temperature, $\rho_{xy}$ is positive and smoothly increases with decreasing temperature.
A sharp minimum with a negative sign appears at $T_{\rm Curie}$ and then $\rho_{xy}$ becomes closer to zero
at lower temperatures.

On the other hand, $\rho_{xy}$ for $H\parallel a$-axis decreases below $T_{\rm Curie}$ with a tiny but sharp minimum at $T_{\rm Curie}$, 
retaining the positive sign.
The sharp kinks at $T_{\rm Curie}$ at $1\,{\rm T}$ in both cases may correspond to the sharp anomaly of susceptibility 
for $H\parallel b$- and $a$-axes at $T_{\rm Curie}$,
which can be defined even at high fields when the field is parallel to the hard axes.

\begin{figure}[tbh]
\begin{center}
\includegraphics[width=1 \hsize,clip]{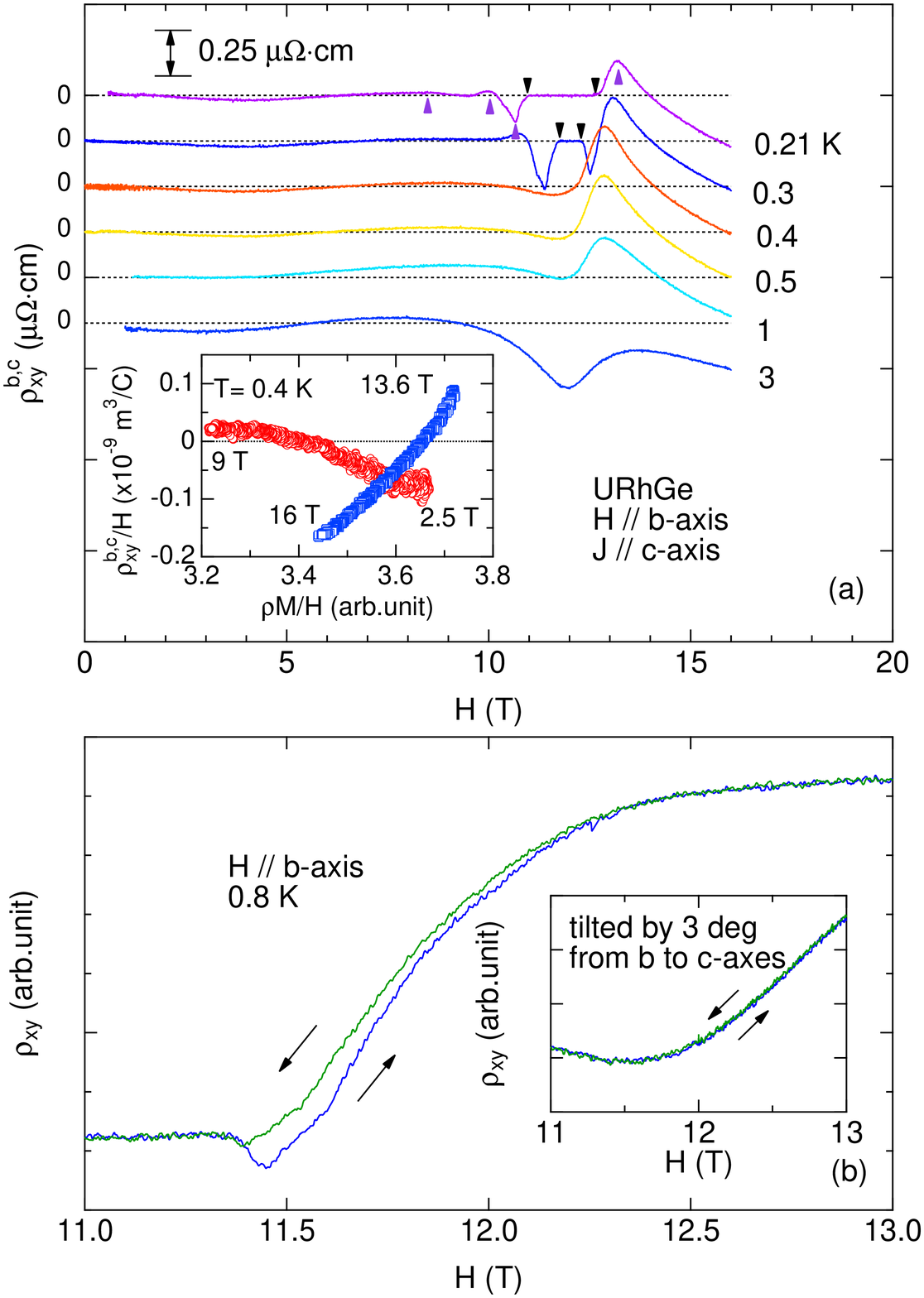}
\end{center}
\caption{(Color online) (a) Field dependences of Hall resistivity for $J\parallel a$- and $c$-axes for $H\parallel b$-axis at $0.4\,{\rm K}$. The spin-reorientation field $H_{\rm R}$ is $12.5\,{\rm T}$.
The inset shows the Hall resistivity at $0.4\,{\rm K}$ in the form of $\rho_{xy}/H$ vs $\rho M/H$ for low fields ($2.5$--$9\,{\rm T}$) and high fields ($13.6$--$16\,{\rm T}$).
(b) Hysteresis of field dependences of Hall resistivity between sweep-up and sweep-down for $H\parallel b$-axis and $J\parallel a$-axis at $0.8\,{\rm K}$.
The inset shows the Hall resistivity when the field is slightly tilted by 3 deg.\ from the $b$-axis to the $c$-axis.}
\label{fig:Hall_Hdep_b}
\end{figure}

Figure~\ref{fig:Hall_Hdep_b}(a) shows the field dependence of the Hall resistivity for $H\parallel b$-axis 
at low temperatures below $3\,{\rm K}$.
At $0.21\,{\rm K}$, $\rho_{xy}$ is almost constant, crossing zero up to $13\,{\rm T}$.
RSC is observed in the field range from $11$ to $12.5\,{\rm T}$, as indicated by small downward arrows in Fig.~\ref{fig:Hall_Hdep_b}(a).
The sharp positive jump at $13\,{\rm T}$ corresponds to the recovery of the normal state after the spin reorientation at $H_{\rm R}\sim 12.5\,{\rm T}$.
With further increasing field, $\rho_{xy}$ rapidly decreases with a sign change from positive to negative.

Fine structures are also found at $0.21\,{\rm K}$, as indicated by small upward arrows in Fig.~\ref{fig:Hall_Hdep_b}(a).
These anomalies are immediately smeared out by increasing the temperature.
This behavior seems to be similar to the results obtained by the thermopower measurements in UCoGe, URu$_2$Si$_2$, and YbRh$_2$Si$_2$, where many anomalies are detected as a function of field only at low temperatures.~\cite{Mal12,Pou13,Pou13_YbRh2Si2,Pfa13}

No RSC is observed at higher temperatures ($T \geq 0.4\,{\rm K}$).
Note that the field dependence of $\rho_{xy}$ with a poor-quality sample 
(RRR $\sim 5$, not shown here) is highly different from that with a high-quality sample (RRR $\sim 40$) shown in Fig.~\ref{fig:Hall_Hdep_b}(a), 
although the anomaly due to the spin reorientation is clearly observed in both cases.
This is most likely due to the large contribution of the anomalous Hall effect, which includes the magnetoresistance.

The first-order transition at $H_{\rm R}$ was clearly detected in Hall resistivity.
Figure~\ref{fig:Hall_Hdep_b}(b) shows the field dependence of $\rho_{xy}$ near $H_{\rm R}$ for $H\parallel b$-axis,
using a different sample with the fine tuning of the field direction by rotating the sample.
The temperature was maintained at $0.8\,{\rm K}$ to avoid any trace of RSC.
A clear hysteresis between up- and down-sweep fields is observed at $0.8\,{\rm K}$,
indicating the first-order transition.
When the field direction is slightly tilted by 3 deg from the $b$ to $c$-axes,
no hysteresis is found within the experimental precision, as shown in the inset of Fig.~\ref{fig:Hall_Hdep_b}(b).
The broad jump of $\rho_{xy}$ is related to the jump of magnetization, as shown in the inset of Fig.~\ref{fig:Mag_phase}(a).

\begin{figure}[tbh]
\begin{center}
\includegraphics[width=1 \hsize,clip]{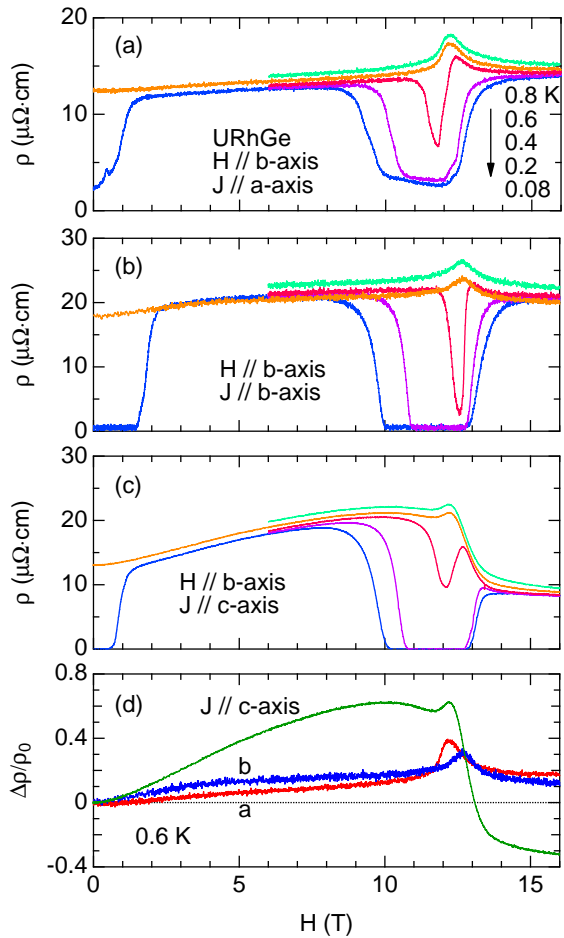}
\end{center}
\caption{(Color online) (a)(b)(c) Magnetoresistances for $H\parallel b$-axis with $J \parallel a$-, $b$- and $c$-axes. Panel (d) shows the magnetoresistances for $H\parallel b$-axis at $0.6\,{\rm K}$ with $J\parallel a$-, $b$-, $c$-axes in the form of $\Delta \rho/\rho_0$ vs $H$.}
\label{fig:MR}
\end{figure}

Figures~\ref{fig:MR}(a)-\ref{fig:MR}(c) show the magnetoresistance for $H\parallel b$-axis with different current directions $J\parallel a$-, $b$-, and $c$-axes for different temperatures.
All data show the SC and RSC, although the magnetoresistance for $J\parallel a$-axis shows only a drop of resistivity instead of zero resistivity because of the sample quality.
The critical fields of SC and RSC slightly differ among $J\parallel a$-, $b$-, and $c$-axes,
because of the sample quality and small misorientation against the field direction within 1 deg.

Above $0.6\,{\rm K}$, all magnetoresistances with different current directions 
show a peak at $H_{\rm R}\sim 12\,{\rm T}$, which corresponds to the spin reorientation.
For $J\parallel a$-axis with transverse configuration, the positive magnetoresistance is observed,
and magnetoresistance shows a slightly higher value above $H_{\rm R}$ than below $H_{\rm R}$,
as we previously reported in Ref.~\citen{Miy08}.
For $J\parallel b$-axis corresponding to the longitudinal configuration, 
the magnetoresistance is positive and almost constant above $5\,{\rm T}$. 
At high fields above $H_{\rm R}$, the magnetoresistance is slightly smaller than that below $H_{\rm R}$.
For $J\parallel c$-axis, the initial positive magnetoresistance changes into the negative magnetoresistance above $H_{\rm R}$
with a large decrease in $\Delta \rho/\rho_0$, as shown in Fig.~\ref{fig:MR}(d)

The response of the magnetoresistance in heavy fermion compounds can have different contributions 
such as the enhancement of the elastic and inelastic resistivity terms
on crossing magnetic instability, valence instability or Fermi surface instability
with feedbacks on $k_{\rm F}$ and $m^\ast$.
Qualitatively, the response of the magnetoresistance is a mark of electronic instability with emerging maxima 
regardless of the current direction for the three configurations.
It is clearly related to the extrapolated enhancement of the $\gamma$-value observed at $H_{\rm R}$.

In $3d$-electron systems,
it is known that the anisotropic magnetoresistance with different current directions is mainly due to 
the anisotropic spin-orbit coupling in the ferromagnets,
where the different densities of states between up and down spins contribute to the different magnetoresistances.~\cite{Cam82}
When the current direction is perpendicular to the direction of the moment, $J \perp M$,
the magnetoresistance decreases,
while the magnetoresistance can increase for $J\parallel M$.
This behavior is also observed in the $5f$-electron system, such as UCoAl~\cite{Mat00}.

In URhGe, the magnetoresistance for $J\parallel c$-axis can be explained by the spin reorientation,
where $J \perp M$ is realized above $H_{\rm R}$, showing the decrease in magnetoresistance,
while at low fields below $H_{\rm R}$, the moment starts to tilt gradually from the $c$ to $b$-axes; thus, 
the behavior of the magnetoresistance is not simple.

To analyze the field dependence of $\rho_{xy}$ for $H\parallel b$-axis,
the data of magnetoresistance in Fig.~\ref{fig:MR}(c) and magnetization data in the inset of Fig.~\ref{fig:Mag_phase}(a) were used,
following Eq.~(\ref{eq1}) with the skew scattering $R_{\rm s}M \propto \rho M$.
The inset of Fig.~\ref{fig:Hall_Hdep_b}(a) shows the plot in the form of  $\rho_{xy}/H$ vs $\rho M/H$
in a low-field range from $2.5$ to $9\,{\rm T}$ and
in a high-field range from $13.8$ to $16\,{\rm T}$.
The extrapolation of $\rho_{xy}/H$ for $\rho M/H \to 0$ corresponds to the Hall coefficient.
As a rough estimate, 
    $R_0 = 2.2 \times 10^{-11}\,{\rm m^3/C}$ at low fields below $H_{\rm R}$
and $R_0 = -1.6\times 10^{-9}\,{\rm m^3/C}$ at high fields above $H_{\rm R}$,
which was extracted from the quadratic extrapolation for $\rho M/H \to 0$.
Using these values, we obtain
$R_0 H      = 0.0055\,{\mu\Omega\!\cdot\!{\rm cm}}$ and 
$R_{\rm s}M =-0.026 \,{\mu\Omega\!\cdot\!{\rm cm}}$ for the low-field FM phase at $2.5\,{\rm T}$.
For high fields in the polarized PM phase,
$R_0 H      = -2.56\,{\mu\Omega\!\cdot\!{\rm cm}}$ and 
$R_{\rm s}M =  2.94\,{\mu\Omega\!\cdot\!{\rm cm}}$ at $16\,{\rm T}$.
Note that the sign of the anomalous Hall effect also changes below/above $H_{\rm R}$.

It is difficult to estimate the precise carrier number; nonetheless
the large difference in $R_0$ between $H < H_{\rm R}$ and $H > H_{\rm R}$
is indicative of Fermi surface reconstruction at $H_{\rm R}$.
Assuming a simple one-band model, 
the Fermi surface is smaller above $H_{\rm R}$ than below $H_{\rm R}$.

However, URhGe is a multiband system; thus, the interpretation of the Hall effect is not simple.
The large mobility, that is, light carrier and long scattering lifetime, mainly
contributes to the normal Hall coefficient.
Thus, complementary experiments, such as thermoelectric power measurements, which are dominantly sensitive for the heavy band, 
or quantum oscillation measurements as a microscopic probe, are required to determine the Fermi surface change more precisely.
However, the present experimental results indicate that at least part of the Fermi surface is
strongly modified at $H_{\rm R}$.

URhGe is a compensated metal with equal carrier numbers of electrons and holes in both the FM and PM states.
The Fermi surfaces in the FM state consist of four different bands,
according to the band structure calculation based on the spin-polarized LAPW method with the 5$f$-itinerant model~\cite{Yamagami}.
The calculated Fermi surfaces in the FM state are highly different from those in the PM state.
The Fermi surfaces in the PM state also consist of four different bands, but the shape of the Fermi surfaces differs from that in the FM state.
Furthermore the total volume of the calculated Fermi surface corresponding to the carrier number  
is larger in the FM state than in the PM state.

In recent ARPES experiments~\cite{Fuj14}, the 5$f$ electron is found to be itinerant.
The change of the electronic structure in the FM state is also found,
although the observed bands are not fully in agreement with those obtained from the calculations.

For $H\parallel b$-axis, $T_{\rm Curie}$ decreases with the field, as shown in Fig.~\ref{fig:Mag_phase}(b).
It is connected with the first order transition at $H_{\rm R}$
with a large change of the sublattice magnetization, as observed in UGe$_2$ from the FM2 phase to the FM1 phase or from the FM1 phase to the PM phase.
When a first-order transition occurs with large change of the FM sublattice magnetization, a marked change of the Fermi surface is generally expected,
in agreement with the present results showing the large change in the Hall coefficient.
In URhGe, the change of the Fermi surface occurs between the low-field PM and FM phases at $T_{\rm Curie}$ on cooling.
By entering into the new phase above $H_{\rm R}$ for $H \parallel b$-axis,
the extrapolation of the magnetization suggests a zero-field FM component $M_0^b\sim 0.1\,\mu_{\rm B}$,
which is far lower than the zero-field FM moment $M_0^c\sim 0.4\,\mu_{\rm B}$ directly obtained for $H\parallel c$-axis, as shown in the inset of Fig.~\ref{fig:Mag_phase}(a).
Thus, the change of the Fermi surface topology at $H_{\rm R}$ can be regarded as an analogous case 
to UGe$_2$ in which the ground state changes either from FM2 ($M_0\sim 1.5\,\mu_{\rm B}$) to FM1 ($M_0\sim 1\,\mu_{\rm B}$), or from FM1 to PM if the component of the magnetic moment for $H\parallel b$-axis is zero ($M_0^b \sim 0$).

In URhGe, the FM structure is collinear. 
The Brillouin zone in the FM state 
is not modified from that in the PM state.
Thus, the change of the Fermi surface at $H_{\rm R}$ is not due to a change of Brillouin zone.
One can consider that the Lifshitz-type transition occurs at $H_{\rm R}$ and modifies the Fermi surfaces.
In SdH experiments, one small pocket Fermi surface ($F=5\times 10^6\,{\rm Oe}$) with heavy cyclotron mass ($m^\ast \sim 20\,m_0$) 
is detected below $H_{\rm R}$,~\cite{Yel11}
which seems to collapse at $H_{\rm R}$. 
This restricted observation suggests the fact 
that the Lifshitz transition has the driving mechanism for RSC.
However, the detected Fermi surface carries only $1.5\,{\%}$ of the total $\gamma$-value. 
It is difficult to explain the large change in $\gamma$-value from $160$ to $220\,{\rm mJ\,K^{-2} mol^{-1}}$ at $H_{\rm R}$, 
as experimentally observed in Ref.~\citen{Har11}.

Our results, which show a large change in the Hall coefficient at $H_{\rm R}$, is consistent with the Fermi surface reconstruction at $H_{\rm R}$
observed by SdH experiments.
It should be noted that the Lifshitz-type transition is not necessarily associated with the first-order transition in general,
because it is basically a ``continuous'' evolution of the Fermi surface due to the Zeeman effect.
In thermopower measurements at low temperatures, many anomalies are often observed
as a function of field, for example in UCoGe, URu$_2$Si$_2$, and YbRh$_2$Si$_2$~\cite{Mal12,Pou13,Pou13_YbRh2Si2,Pfa13}.
The other sign changes in $\rho_{xy}$ at low temperatures below $H_{\rm R}$ in Fig.~\ref{fig:Hall_Hdep_b}(a) may
indicate the precursor effect of the Fermi surface evolution with the field.

Our experiments confirm that the Fermi surface is strongly modified
when the ordered moment is changed.
For comparison, the Hall resistivities of three ferromagnetic superconductors are shown in Fig.~\ref{fig:Hall_compare}.
As shown in Fig.~\ref{fig:Hall_compare}(a), the Hall resistivity in the ferromagnetic superconductor UGe$_2$
changes markedly at $H_{\rm c}$ which separates the PM and the FM1 states through a first-order transition,
indicating a marked change of the Fermi surface.

In UCoGe, surprisingly, no clear anomaly is observed in the Hall resistivity for $H \parallel b$-axis~\cite{Aok14_SCES},
however, the thermopower measurements detect the anomaly at approximately $12\,{\rm T}$, implying the Fermi surface change.~\cite{Mal12}
In the magnetization measurements of UCoGe, no clear anomaly was found so far for $H\parallel b$-axis,
because the ordered moment ($m_0\simeq 0.05\,\mu_{\rm B}$) is one order of magnitude smaller than that in URhGe,
and the ferromagnetic transition is not very clearly detected compared with that in UGe$_2$ or URhGe.
Furthermore, the initial slope of magnetization for $H \parallel b$-axis, namely, $dM/dH |_{H\to 0}$ is not very 
large compared with that for $H\parallel c$-axis~\cite{Huy08,Kna12}, which is not favorable for spin reorientation.
\begin{fullfigure}[tbh]
\begin{center}
\includegraphics[width=1 \hsize,clip]{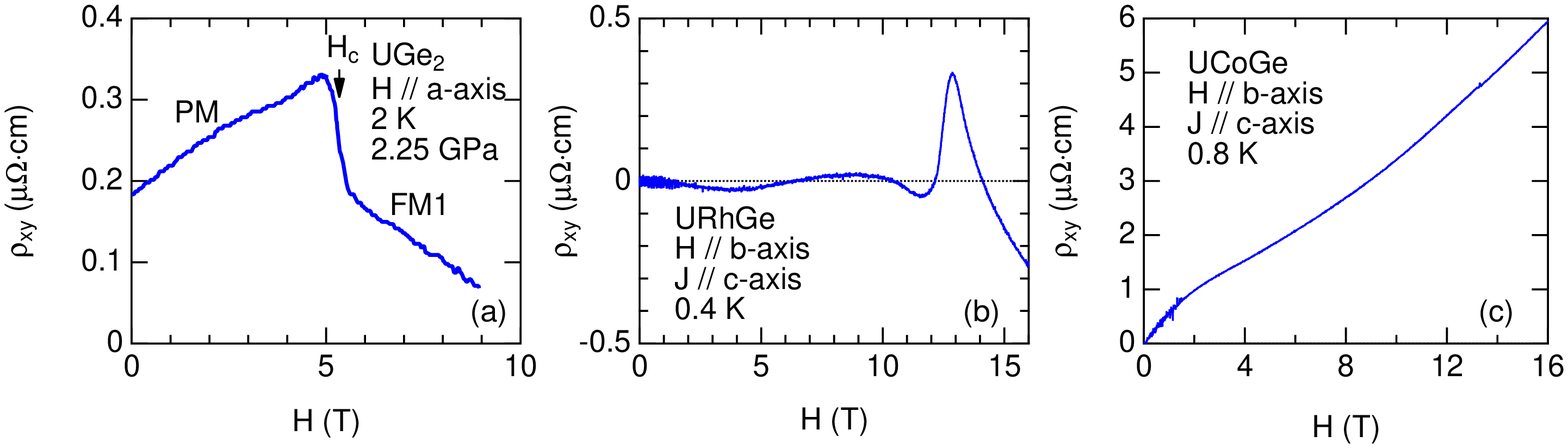}
\end{center}
\caption{(Color online) Field dependences of Hall resistivity in (a) UGe$_2$~\cite{Kot11}, (b) URhGe and (c) UCoGe~\cite{Aok14_SCES}.}
\label{fig:Hall_compare}
\end{fullfigure}

\section{Summary}
We measured the Hall resistivity of URhGe using high-quality single crystals.
The Fermi surface change between PM and FM at low fields was observed from the change in the Hall coefficient.
The Fermi surface further changes when a field is applied along the $b$-axis across the 
spin-reorientation field $H_{\rm R}$.
This change is most likely explained by the Lifshitz-type transition associated with magnetic instabilities
related to the marked change of the FM sublattice magnetization.

\section*{Acknowledgements}
We thank S. Araki, S. Fujimori, H. Harima, L. Malone, K. Miyake, A. Pourret, and H. Yamagami for useful discussions.
This work was supported by ERC starting grant (NewHeavyFermion), French ANR project (CORMAT, SINUS, DELICE), KAKENHI, ICC-IMR, and REIMEI.


\end{document}